\documentclass{llncs}
\usepackage{graphicx}
\usepackage{colortbl}
\definecolor{myblue}{rgb}{0.85,0.85,0.85}

\begin{document}
\mainmatter
\title{Towards a Swiss National Research Infrastructure}
\author{Peter Kunszt\inst{1} \and Sergio Maffioletti\inst{2}
Dean Flanders\inst{3} \and Markus Eurich\inst{1} \and Eryk Schiller\inst{8}
\and Thomas Michael Bohnert\inst{4} Andy Edmonds\inst{4} \and Heinz
Stockinger\inst{5} \and Almerima Jamakovic-Kapic\inst{6} \and Sigve
Haug\inst{6} \and Placi Flury\inst{7} \and Simon Leinen\inst{7}}
\authorrunning{Peter Kunszt et.al.}
\tocauthor{Peter Kunszt, Andy Edmonds, Dean Flanders, Placi Flury,
  Sigve Haug, Almerima Jamakovic-Kapic, Sergio Maffioletti and Heinz
  Stockinger}
%
\institute{ETH Z\"urich, Clausiusstrasse 45, 8092 Zurich, Switzerland,\\
\and
Universit\"{a}t Z\"urich,
Grid Computing Competence Center,\\
Winterthurerstrasse 190, 8057 Zurich, Switzerland,
\and
FMI, Maulbeerstrasse 66, 4058 Basel, Switzerland
\and
ZHAW, Obere Kirchgasse 2, 8400 Winterthur, Switzerland,\\
\and
SIB Vital-IT, Quartier Sorge, 
1015 Lausanne, Switzerland
\and
University of Bern, 
3012 Bern, Switzerland
\and
SWITCH, 
Werdstrasse 2, 8004 Zurich, Switzerland
\and
University of Neuch\^atel, 
Rue Emile-Argand 11, 2009 Neuch\^atel, Switzerland
}

\maketitle     

\begin{abstract}
In this position paper, we describe the current status and plans for a
Swiss National Research Infrastructure. Swiss academic and research
institutions are very autonomous. While being loosely coupled, they do
not rely on any centralized management entities. A
coordinated national research infrastructure can only be established
by federating the local resources of the
individual institutions. We discuss current efforts and business models
for a federated infrastructure.

\keywords{grid, cloud, distributed computing, federation,
  funding}

\end{abstract}
%

\section{Introduction and Overview}
Switzerland, as a country, is organized in a very decentralized
manner, which has a deep impact on the organization of research and
education.  At the federal level, there is the
national supercomputing center, CSCS in Lugano, the national research
network SWITCH and several institutes of research and technology of the ETH Domain. At
the cantonal level, there are ten universities, nine universities of
applied sciences, and several specialized institutions for research and
education. The organizational and funding
structures are very heterogeneus. Today, infrastructure provisioning happens
exclusively on the institutional level. We believe that the scientific
research community, as well as resource providers, may greatly benefit
from a coordinated, federated national research infrastructure.


There exist various architectures of federated
clouds~\cite{morenovozmediano12} such as hybrid, broker-based,
aggregated, and multi-tiered. The hybrid architecture allows a private
cloud to cooperate with a public one by using specific drivers for cloud bursting.
For example,
OpenNebula~\cite{sotomayor09,morenovozmediano11} can cooperate with
Amazon (Amazon EC2 driver), and Eucalyptus~\cite{eucalyptus}.
StratusLab~\cite{stratuslab}, an OpenNebula-based cloud
infrastructure, provides a toolkit to integrate the
most recent cloud management technologies.  In the broker based model,
the broker coordinates the resources of several public clouds. There are a few
online broker sites such as BonFire (www.bonfire-project.eu), Open
Cirrus (opencirrus.org), and FutureGrid (futuregrid.org).

The Reservoir system~\cite{rochwerger11} provides an infrastructure
that may automatically deploy tasks of one organization in
the clouds of other partners having spare capacities. 
A multi-tiered architecture is an interesting option for large
companies with geographically distributed resources. Such resources
are usually strongly coupled as they serve as sub-services
of the same entity. However, we believe that none of these models serve 
our specific needs adequately for Swiss higher education and research.

There are several examples of past Swiss national initiatives working
towards the goal of setting up services to address the needs of several
scientific user communities. The Swiss Multi-Science Computing Grid
project~\cite{smscg} provided services mainly to the Swiss High Energy
Physics community and enabled that community to participate in the
European Grid Initiative (EGI). In the life sciences domain, the SyBIT
project~\cite{sybit} of the Swiss Initiative for Systems Biology, SystemsX.ch, 
makes use of the computing and storage infrastructure of
all participating institutions. Infrastructure is not
organized in a coherent manner, as SyBIT is focusing on bioinformatics 
support, not infrastructure.

The funding for these efforts is project-based, and is aimed to create
infrastructure and services in tight cooperation with a user
community. However, 
it is currently unclear what will happen to the
services, infrastructure and all the other additional value created
(mostly in the form of specialized know-how) after their funding
expires. The user communities still require services beyond their
lifetime.

An example of how long-term sustained funding can be set up, is given
by the Vital-IT group~\cite{vitalit} of the Swiss Institute of
Bioinformatics SIB. Vital-IT provides services for bioinformatics in life
sciences, coordinating bioinformatics infrastructure and resources for
several universities in the Western part of Switzerland. Vital-IT is
funded by the participating universities, through federal funds 
and to a large extent by second
and third parties (industrial and academic research grants). 
Researchers are advised to request these charges as part of their project grant
proposals as 'bioinformatics consumables'.

Very recently, the Swiss government has initiated a project to work
out a strategy for the development of a sustainable set of services
for research, education and digital archiving communities.  The
individual sustainability models for the services developed are to be
established in this project.

In this paper, we describe the current Swiss Academic Compute Cloud
project that we use to prototype a federated Swiss infrastructure for
research computing.  We also summarize our current ideas about
necessary policy and accounting models that have to be established to
allow for a sustainable distributed Swiss research infrastructure. 
This should solve the remaining open issue of sustainable operation.

\subsection{Cloud Definitions}
We believe that it is necessary to specify our own definition of
'cloud' since the term has been used in many, sometimes conflicting
contexts.

We expect the following properties from a service with
a 'cloud' attribute:
\begin{description}
\item[Self serviced] A consumer has immediate access to resources and
 can unilaterally access computing capabilities, such as server
 time and network storage.
\item[On-demand] As needed, at the time when needed, with the
  possibility of automatic provisioning. 
  No need to do the full investment planning upfront.
\item[Cost transparent] Accounting of actual usage is completely
  transparent to both the user and service provider, which is measured
  in meaningful terms.
\item[Elastic, scalable] Capabilities can be elastically provisioned
  and released, in some cases automatically, to scale rapidly up and
 down, matching demands. 
\item[Multi-tenant] The provider’s computing resources are pooled to
  serve multiple consumers, with resources dynamically assigned and
  reassigned according to consumer demands. Ideally, for the customers,
 other tenants are invisible.
\item[Programmable Services] The services expose a public programmable
  API that can be used to drive any aspect of the service
  programmatically.
\end{description}

These properties are technology-agnostic; the way we define
cloud services has a consequence on how we expect the services to be
exposed to the consumer and what kind of business model and usage
policy is in place.

We also use the terms Infrastructure as a Service (IaaS), Platform as a
Service (PaaS) and Software as a Service (SaaS) to define different
fundamental types of services offered to researchers and
educators. All these service types are considered to be part of
the cloud ecosystem, they can be built on top of each other or
directly on the underlying infrastructure.

\begin{figure}[htbp]
    \centering
        \includegraphics[width=1.00\textwidth]{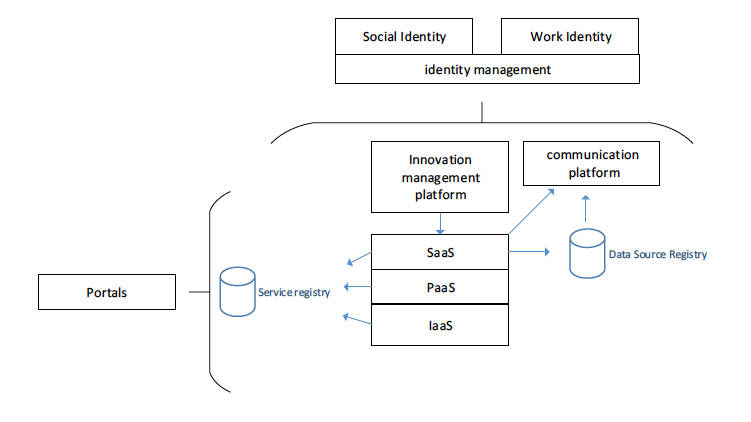}
    \caption{Architecture of the digital ecosystem}
    \label{fig:arch}
\end{figure}

\section{Ecosystem Architecture}

The development of a robust {\it digital ecosystem} is key to maintain competitiveness of Swiss research. Researchers need easy
access to resources for their work and to maintain collaborations across
institutional boundaries. 
We propose the establishment of an open informatics platform, which 
can be used to create the digital ecosystem by using clouds as 
an innovation enablement technology for sharing all types of resources.

The architecture of the digital ecosystem is based on cloud services,
but at a very high level, it can be divided into the following elements
(see Figure \ref{fig:arch}):
\begin{itemize}
\item Identity management (authentication of social and work identities)
\item Social platform (information of interest to the researcher)
\item Ideation platform (process of developing new services)
\item Service Portals (access to resources)
\end{itemize}

This will provide a platform for researchers, research institutions,
societies, and associations to rely on cloud services so that they 
can focus their energies in areas of value creation. One such
example is FASEB~\cite{faseb} (Federation of American Societies in
Experimental Biology), which supports individual researchers and
societies with tasks common to all of them.

In terms of {\it identity management}, every member of an institution
of higher education in Switzerland has already an identity in a
Shibboleth-based federated identity management infrastructure~\cite{switchAAI}.
However, this infrastructure needs to be extended with additional
services like inter-institution group management, to meet the needs of
the full ecosystem. 

{\it Social} and {\it ideation platforms} are instruments aimed to
channel researcher requests, needs and ideas into a discussion network, 
in which problems (expressed in terms of challenges) are evaluated and
discussed, and where solutions are agreed, designed, implemented, and
then added to the service portals as part of the available service
portfolio.

{\it Service portals} are a generic model to introduce new technology that
can be offered and adapted to meet the needs of a variety of researchers.  
The services introduced will be based on use-cases delivered by
researchers or by the ideation process.



\section{The Swiss Academic Compute Cloud Project}
The Swiss Academic Compute Cloud Project (SwissACC) has been set
up to keep and extend the currently established Swiss-wide
computational science platform composed of resources and services of
various types (local clusters, grid and cloud infrastructures) as well
as the excellent know-how for user and application support. Research
communities can profit from this platform either to address their
computational needs, or to make use of the platform for collaboration
purposes in national and international projects. It provides IaaS,
PaaS and SaaS services to communities and helps to build their own
service portals or tools.

The project builds upon previous initiatives at the national level
such as the Swiss Multi Science Computing Grid (SMSCG)~\cite{smscg},
the Academic Compute Cloud Provisioning and Usage~\cite{acadcloud}, 
VM-MAD~\cite{vmmad}, GridCertLib~\cite{gridcertlib}, RS-NAS~\cite{rsnas};
and international projects like EGI~\cite{egi}, 
Chemomentum~\cite{chemomentum}, and PRACE~\cite{prace}.

The primary goal of SwissACC is to sustain and keep together the
research communities that have been brought to distributed
infrastructures in the previous Grid and Cloud projects.
The secondary goal is to increase the number of supported research
communities. A large community can profit from mutual benefits due 
to economies of scale as they gain more experience and receive more
feedback, improving the overall quality of the provided services.

The final goal is to provide a sound basis for decision making on
infrastructure models, like in-house vs. outsourcing, and on
technologies, like running a cluster, a virtualized infrastructure, or
going to a public cloud provider.

The project is composed of infrastructure providers, service
providers and user communities that have very close and regular interaction
with each other. This allows the project to continuously improve the services
and to focus on items that users really need and use. The user support
is set up using a collaborative distributed support model, helping
scientists in bringing their small and large-scale data analysis
pipelines to a flexible cloud-like platform that can easily be shaped
to accommodate their needs. 

The current infrastructure supports user communities who already solve
research problems with help of computational
processing. Virtualization techniques from e.g. the VM-MAD~\cite{vmmad}
and AppPot~\cite{app-pot} projects and a powerful toolbox for job
management on heterogeneous infrastructures (GC3Pie~\cite{gc3pie})
allows us to set up a standardized procedure for enabling new
research communities. These technologies have allowed us to start
supporting new users and communities with relatively little
effort. However, it is not enough to just virtualize a certain
application; there have to be several middleware layers to assure
that this application can scale, e.g. make use of necessary infrastructure
services like storage, and is easy to use by the end user.

The SwissACC project currently supports over 20
applications~\cite{smscg-app} from the domains of life sciences, earth
sciences, economics, computer sciences, engineering, cryptography and
physics. Table~\ref{table1} lists the applications supported, which we
intend to extend throughout the duration of the project.

\begin{table}[!hb]
\begin{center}
  \begin{tabular}[ht]{|l|p{2.8cm}p{4.0cm}p{2mm}p{2.6cm}|}
    \rowcolor{myblue}

    \hline
    Institution&Applications&Characteristics& &Research Field\\[2pt]
    \hline

IMSB/ETHZ & Rosetta,\newline TPP,\newline HCS pipeline & 2-3 users,
152k jobs, 300k walltime hours during last 12 months & & life sciences
/\newline proteomics\\ \hline
IBF/UZH & gpremium & 1 user, 916k jobs, 238k walltime hours, (1.5.2011
-- 1.7.2012) & & Economics /\newline financial models\\ \hline
GEO/UZH & GEOTop & 4-8 users, 37k jobs with 11k walltime hours since
June 2012 & & earth sciences\\ \hline
UniBE/UniNE & A4-Mesh & 15 users, real-time data collected since June
2013 & & environmental research /\newline hydrogeological
modeling\\ \hline
Lacal/EPFL & gcrypto & 1-4 users, 16k jobs, 32k walltime hours
during last 6 months & & cryptography\\ \hline
UNIL/Vital-IT & Selectome & 1 production user. 71k jobs, 190k
walltime hours during last 12 months& &Life-sciences/\newline
phylogeny\\ \hline
UniGE & MetaPIGA & execution model still to be defined. & &
Life-sciences /\newline
engineering\\ \hline
LHEP/UniBe & ATLAS & Several (international) users, ca 27k
jobs/month with 239k walltime hours& &High Energy \newline
Physics\\ \hline
WSL/SLF& CATS,\newline Alpine3D,\newline SwissEx& CATS: 1 user,
1month, 2300 jobs, 38k walltime hours SwissEx: cyclic analysis (30'
frequency) processing data from 3 IMIS stations
& &earth sciences/\newline climate
modeling\\ \hline
\end{tabular}
\caption{Applications on SwissACC}\label{table1}
\end{center}
\end{table}


\subsection{Infrastructure for SwissACC}

The SwissACC project has currently access to 5 OpenStack cloud
installations at the ETH Zurich, the
University of Zurich, SWITCH, the Zurich University of Applied
Sciences and the HES-SO site in Geneva. 
These clouds are relatively small: they range from 100 to
400 CPU cores each, but with relatively good memory and storage.

The project also has access to the Swiss National Grid infrastructure, 
an aggregation of relatively
large clusters, as this is part of the LHC Computing Grid
infrastructure. 

The OpenStack versions of the various installations are not identical.
The coordination does not involve upgrades
being rolled out simultaneously across all sites, and we believe it is
not necessary to aim for such strong coordination as long as
interoperability can be maintained. 
The OpenStack distributions being used are also not identical. Some are
based on Ubuntu, others on Red Hat, Scientific Linux or Fedora.

The choice of OpenStack as a reference cloud software stack has
emerged as an evaluation done in a predecessor project, the Academic
Cloud Provisioning and Usage project~\cite{acadcloud}. There, 
the ETH Zurich and the University of Zurich evaluated the
available cloud solutions and set up several pilot cloud installations.
OpenStack has been evaluated against other public open cloud
stacks (like OpenNebula, Eucalyptus,
CloudStack~\cite{cloudstack}) as well as commercial cloud solutions
(Flexiant, StackOps, HP CloudSystem Matrix, etc.).

Commercial cloud solutions are usually very mature, with several
layers of well-tested software built into themselves, leveraging these
companies' experience in operating large-scale automated
infrastructures. These layers make them very complex and not easy to
maintain without proper training. Therefore, such systems should
be optimally bought directly from the company including expert support,
preferably by an on-site full-time specialist.

Lacking the funding for such a system and the expert, we chose
OpenStack as the open cloud software stack, since currently it has the
largest community and commercial support. However, 
OpenStack is not yet fully mature as a production
system. It is already usable and there are very large
installations running it (for example at CERN), but the expectations
of what is possible and what the challenges are need to be set
correctly, also towards the users.  We have exposed a lot of stability and
security issues, that are being addressed in newer releases. For example, the
accounting and quota management functionality is still largely in
development. Our recommendation to our partner resource providers is
therefore to slowly ramp-up cloud services offered to the researchers,
with proper training and expectation management, initially being
mostly for testing and educational purposes. With the right
high-level middleware like GC3Pie~\cite{gc3pie}, the free OpenStack
solution is already adequate for many use-cases. For more reliable
services and more complete installations, a commercial
OpenStack-compliant solution might be considered, as offered by
Rackspace, Red Hat, etc.




In terms of cloud coordination for technology and adoption, we
have initiated a dedicated interest group for cloud computing in the
Swiss Informatics Society~\cite{sigsi}.

\section{Cloud Charging Models for Federated Infrastructures}

In order to exploit economies of scale, i.e., the consolidation of
resources, the institutions need to be able to provide services to 
each other at a cost. 
However, there is currently no model and in fact no sufficient legal
basis that would allow the institutions to charge each other for their
respective incurring service costs. We are therefore proposing several
cloud-like business models for the provisioning of services by
higher-education centers, allowing for many different charging models
for them to choose from, in the hope of starting a process that will
eventually allow a much tighter cooperation than is possible today.

Also as part of the Academic Cloud Provisioning and Usage project~\cite{acadcloud},
we tried to gain insights
into service consumers' and providers' needs as a basis
for a pricing strategy of cloud-based services and for future
decision-making. Information was gathered by interviewing research
groups currently using central computing services.

The interviews revealed that academic service consumers currently perceive
cloud services as a playground for testing, experimenting, and training
students. However to date, most academic service consumers do not
fully perceive the added value that the cloud computing model can
provide.
Some groups say they would use cloud services if they were
available at a competitive cost. Currently, these cloud-based services
are not necessary due to existing private infrastructures or service providers.
These service providers do, however, see several advantages that a 
cloud model could provide in the future, such as flexibility,
provisioning of additional  services, time to service, self-serving
aspects and increased automation, elasticity, and a more balanced
workload.

There are three basic pricing schemes that are acceptable for
academic consumers of cloud-based services: 
\begin{description}
\item['Pay per use'] Service consumers are charged a fee according to
  the time and volume of a computing service that has been
  consumed. 
\item[Subscription]  The service consumer pays a fee on a regular 
  basis for the usage of a service.
  Subscriptions allow services to be sold as packages.
\item['Pay for a share'] In this approach service consumers buy a
  share in order to get a corresponding amount of service.
\end{description}
It is of course possible to offer
several of the three models above simultaneously or in a mixed form,
for example adjusting the pay-per-use pricing on a pre-subscription
volume, as it is already done by Amazon.  

The interviews revealed that academic service consumers care little
about pricing strategies. They want to focus on science and
research. Foremost, they just want services that work. They expect an
easy approach and want the university to clear all of the obstacles
from their paths. From their perspective, the 'per pay use' and
subscription approach can only work if there is a way to get funding
for it.

From the service providers' perspective, there is no pricing scheme
that outperforms any of the other schemes. Each of the three pricing
schemes has certain advantages and disadvantages. The usefulness of
one approach can depend on the length of time a given service is needed. ‘Pay per use’
qualifies best for a short-term service demand that lasts no more than
three months or as an addition for peak usage on top of a subscription
or share. The subscription scheme might be a good solution for mid- to
long-term commitments. The ‘pay for a share’ approach may be most
suitable for long-term commitments, i.e., for more than two years. In
our definition of cloud we of course favor the pay-per-use model as it
fits the operational-expense-only model best. However, until the
funding obstacles for such a model are overcome (both on the provider
and the consumer side), we need the other schemes at least during a
transition phase, the length of which is to be determined.

Finally, many academic service consumers do not know much about cloud
computing and in fact do not care much about what computing resources
are used. They just need some powerful computing services in order to
do their research. Therefore, the pricing scheme should be simple,
yet fair and transparent. In this context, subsidization
can play an important role to clear inconveniences out of the service
consumers' way.
The institutions might set up a subsidization scheme, providing the
researchers with free access to resources, or other means for the researchers to pay for
cloud-based services. 

To the funding agencies, we propose to introduce an 'informatics
consumable' concept based on the experience of the Vital-IT competence
center. The idea is that projects simply put in a line item called
'informatics consumables', requesting funding to be spent on the
available research computing infrastructure to meet their
computing and storage needs. 
Obviously, the evaluation process applied by the funding agencies should take the
amount of requested informatics consumables into consideration and the
process should check back with the infrastructure providers to find
out whether the request is technically sound and whether the requested
consumables are adequate.  In the Vital-IT model, such consumables
also apply to user support services and to standardized data analysis
services performed by bioinformaticians at the competence center,
providing expertise to projects in this domain that do not have the
project partners to perform a standardized analysis of their data.

\section{Summary and Outlook}

We have presented the current efforts to maintain a federated research
computing infrastructure in Switzerland, with details on our thoughts
for sustainability and funding models.

We are working towards a sustained national infrastructure model based
on federation concepts where researchers can easily gain access to the
necessary resources and support persons through local contacts at
their individual institutions. Sustainable funding models have been
presented but they still need to be refined, accepted and put in place
by funding bodies and institutions to have a lasting,
sustained effect on the Swiss research landscape.

Given the high expectations that cloud technologies set, there are
certain risks that we need to be aware of and address properly. Based
on our experience with federation in the context of Grid technologies,
we know that it is very important to get actual scientific users on
board early, to set expectations correctly and to immediately use new
technologies and services.

Our goal is to advance scientific research by lowering the barriers to
usage and adoption of information technologies in general. In an ideal
scenario, a scientist has the choice to select from a large array of
easy to use tools and services to perform his or her research. 
He or she should also be able to easily publish new scientific
services based on their individual research or to contribute to
the existing community of tools and services. Such frameworks and services
need to be intuitive, resilient to failures, provide meaningful
error messages, and integrate with social media to interact with
other scientists and service supporters in the right context.

%
%

\end{document}